# Capacitive microwave sensor for toxic vapor detection in polluted environments

P. Bahoumina[1*§], H. Hallil[1°§], K. Pieper[1], J.L. Lachaud[1], D. Rebière[1], C. Dejous[1§]

[1]Univ. Bordeaux, Bordeaux INP, CNRS, IMS UMR 5218
F-33405 Talence, France

A. Abdelghani[2], K. Frigui[2], S. Bila[2], D. Baillargeat[2]

[2]Univ. Limoges, CNRS, XLIM UMR 7252
F-87060 Limoges, France

Q. Zhang[3], P. Coquet[3,4]

[3]NTU, THALES, CNRS, CINTRA UMI 3288
Singapore 637553, Singapore

E. Pichonat[4], H. Happy[4]

[4]Univ. Lille, CNRS, IEMN UMR 8520
F-59652 Villeneuve d'Ascq, France

(*)prince.bahoumina@ims-bordeaux.fr;
(°)hamida.hallil-abbas@u-bordeaux.fr,
(§)IEEE Member

*Abstract*—This paper presents an inkjet printing capacitive microwave sensor for toxic vapor detection. The designed sensors were presented and fabricated with success. The experiments show sensitivity to ethanol vapor according to the S parameters. It is equal to 0.9 kHz/ppm and 1.3 kHz/ppm for the sensors based on 5 and 50 sensitive layers respectively. This sensor will be integrated into real-time multi-sensing platforms adaptable for the Internet of Things (IoT).

## I. Introduction

Toxic compounds in the polluted environments involve several particles and mixtures of complex gases, such as VOCs. The European Union (EU), the International Agency for Research on Cancer (IARC) and the Institut National de Recherche et de Sécurité (INRS) provide information on the carcinogenic risk of listed chemical substances in the workplace. For example, the occupational exposure limit value for ethanol is 500 ppm in Germany and 1000 ppm in France and the USA [1]. Powerful systems that allow the real-time detection of these chemical agents in the form of prevention tools to alert the people concerned are therefore urgently needed. Currently, most of the commercially available sensors are based on conductivity transduction using metal oxides as a sensitive layer. However, such sensors often operate at high temperatures. On the contrary, electromagnetic transducers can operate at room temperature [2], which is just one of their advantages. In fact, they have been selected because of various advantages, as follows. Besides being a passive device, they could possibly function wirelessly. They are appropriate for networking and communicating operation with a high-frequency working thread, usable for real-time detection and they provide an exploitable information directly. In addition, due to its planar structure, the device can be manufactured on flexible substrate by low cost inkjet printing technology [3]. To increase the sensitivity, several parts of the active channel are covered with a sensitive layer. In this study the sensitive material is based on poly (3,4-ethylenedioxythiophene) polystyrene sulfonate and multi walled carbon nanotubes (PEDOT:PSS-MWCNTs). The designed and fabricated sensors are presented. Electrical measurements of two devices in a frequency range up to 6 GHz as well as the characterization under ethanol vapor are performed.

## II. Theoretical Study

### A. Design and analytical model

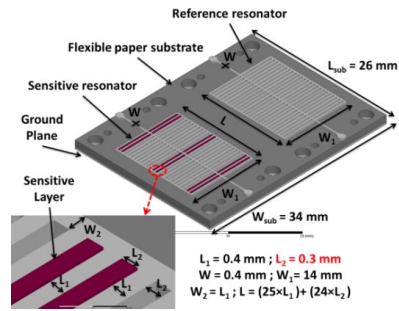

Fig. 1. Dimensions and geometry of sensor based on two resonators

The sensor pattern and dimensions are presented in Fig. 1, with two bandpass resonators to achieve differential detection. A bare resonator is considered as reference channel, while the other resonator is the sensitive channel functionalized with a sensitive material to target the toxic vapor. Each channel consists of two parallel capacitive patterns of 50 electrodes each, with a gap (L2) of 300 μm between two successive electrodes. The sensor operating principle was studied beforehand [2], [4]. Each resonator acts as a bandpass filter with several resonant frequency modes and is dimensioned to produce a first mode at around 2.4 GHz. The resonant frequencies ($f_{rn}$) are defined by:

$$f_{rn} = \frac{n \times c}{2 \times (W_1 + L - W) \times \sqrt{\varepsilon_{eff}}}$$

Where: L and $W_1$ are the real length and width of the resonator, W is the width of the access feed line, c is the velocity of light in free space ($3.10^8$ m/s) and $\varepsilon_{eff}$ is the effective permittivity.

*B. Simulation results*

The HFSS$^{TM}$ finite element tool is used for the design and simulation of the sensor. All simulations are performed in the frequency range from 1 to 6 GHz. In this study, we used a photo paper substrate with the measured electrical properties as a function of the frequency, given in Table I. These values allow us to activate the "Set Frequency Dependency Material" option of the simulation software and to use the "Multipole Debye Model Input" in order to consider the paper properties according to the frequency.

TABLE I. EXPERIMENTAL PAPIER PARAMETERS INTEGRATED IN SIMULATION SOFTWARE.

| Frequency (GHz) | Relative Permittivity | Dielectric Loss Tangent |
|---|---|---|
| 2.45 | 3.08 | 0.1270 |
| 4.70 | 3.05 | 0.0961 |
| 10.00 | 2.79 | 0.0170 |
| 16.00 | 2.76 | 0.0733 |

In actual devices, commercial nanocomposite materials have been used. The silver metallization of the bare resonators is printed from an ink based on silver nanoparticles while the additional sensitive layer is printed from an ink based on a composite polymer (PEDOT: PSS - MWCNT). All measured material parameters were presented previously [4], [5]. The reference resonator can be used to compensate the variations not related to the sensitive material. Fig. 2 shows the S parameters resulting from the simulation of each channel. The first and second resonant mode of the reference and sensitive resonators are recorded around 2.414 and 4.72 GHz. The simulation of the presence of 5 and 50 layers of sensitive material induces a frequency shift close to -111 and +47 MHz for the first modes, and -134 and -115 MHz for the second modes of each number of sensitive layers, respectively. The distribution of the electric field E at both resonant modes of the reference channel is shown in Fig. 3. It puts to evidence areas with more intense E fields. In these areas, 12 bands, in the 300μm gaps of the measurement channel, were selected to place the 5 and 50 sensitive layers (see Fig. 1), in order to emphasize the interactions with the E field and thus, a capacitive disturbance effect.

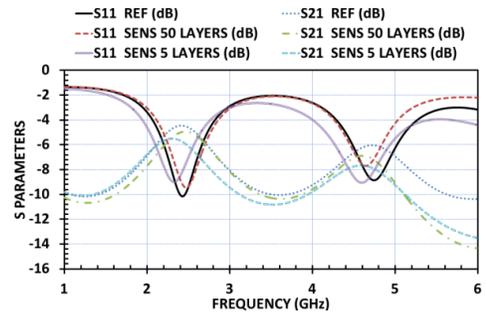

Fig. 2. S parameters simulation results

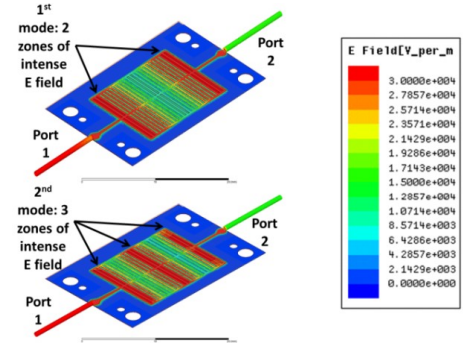

Fig. 3. Distribution of the electric field E of the reference resonator for the first and second modes

III. EXPERIMENTAL STUDY AND DISCUSSION

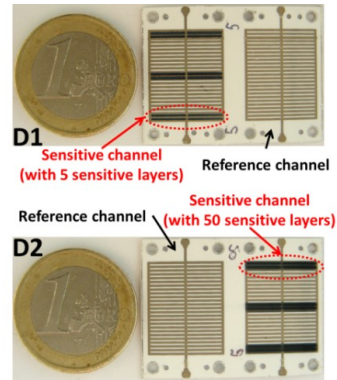

Fig. 4. Manufactured devices (D1 and D2)

The device was manufactured with commercial materials for reduced costs and enhanced reproducibility. We used a Dimatix inkjet printer (2800 series), Metalon® JSB 25 HV silver ink as metallization for the geometry of each resonator and Poly-Ink HC ink (PEDOT:PSS-MWCNTs) as sensitive layer. The substrate is Epson photo paper with a thickness of 260 μm. Two devices were fabricated, as presented in Fig. 4, the first with 5 sensitive layers (D1), the second with 50 layers (D2) in order to improve the sensitivity of the sensor.

*A. Electrical characterization*

The experimental study is made with the Anritsu MS2026B vector network analyzer (VNA) in the frequency range from 1 to 6 GHz. The electrical characterization results of the reference and the sensitive channels of both devices are illustrated in Fig. 5. Both resonators of the devices present two modes in the considered frequency range. The first modes were registered around 2.7 GHz and the second

Research supported by the French National Research Agency under ANR-13-BS03-0010 project.

modes around 5.4 GHz for D1 and D2. The general behavior is consistent with the simulation results presented in part II.

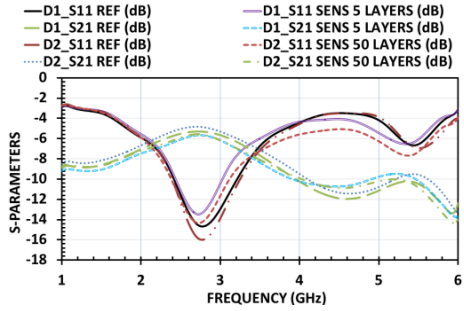

Fig. 5. S parameters of manufactured devices (D1 and D2)

*B. Ethanol vapor detection*

In this section, we are focusing on the characterization around the first resonant modes of the $S_{11}$ parameters of the resonators of each device, in the reduced frequency range from 2.1 to 3.1 GHz for vapor detection with the same vector network analyzer, calibrated with 4001 points.

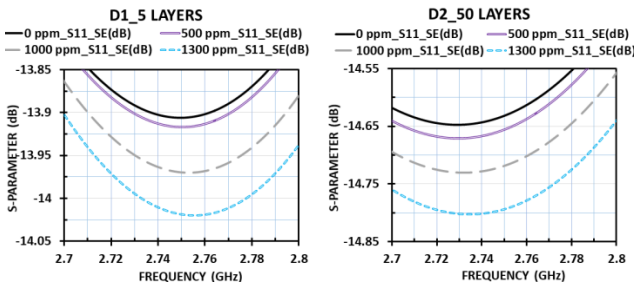

Fig. 6. $S_{11}$ of D1 and D2 for each concentration of ethanol vapor

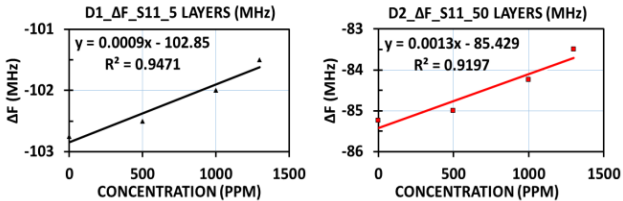

Fig. 7. Differential frequency variation as a function of different concentrations of ethanol vapor

The experimental detection configuration is mainly based on a vapor generator. The target vapor is transported by nitrogen as a carrier gas [4]. Ethanol was used as the target vapor, with concentrations of 0, 500, 1000 and 1300 ppm. The device was exposed to the vapor for 10 minutes for each concentration at room temperature. The 0 ppm concentration corresponds to nitrogen exposure only. Fig. 6 illustrates the effect of the ethanol vapor on the resonant frequencies of the resonators according to the reflection parameters ($S_{11}$). The resonant frequencies of the reference (FrS11r) and sensitive (FrS11s) resonators increase for both devices under exposure to the target vapor. This is in accordance with a decrease of the permittivity of the environmental medium as well as of the sensitive layer as shown in the relationship 1 in part II. For a clearer understanding, the results can be represented in differential detection mode, with ΔF_S11 obtained by subtracting the response of the reference from that of the sensitive channel as given in Table II.

TABLE II. RESONANT FREQUENCY VALUES ACCORDING TO DIFFERENT CONCENTRATIONS OF ETHANOL VAPOR

| C (ppm) | D1_FrS11r (MHz) | D1_FrS11s (MHz) | D1_ΔF_S11 | D2_FrS11r (MHz) | D2_FrS11s (MHz) | D2_ΔF_S11 |
|---|---|---|---|---|---|---|
| 0 | 2.85200 | 2.74950 | -102.75 | 2.81425 | 2.72900 | -85.25 |
| 500 | 2.85300 | 2.75025 | -102.50 | 2.81450 | 2.72950 | -85.00 |
| 1000 | 2.85500 | 2.75300 | -102.00 | 2.81625 | 2.73200 | -84.25 |
| 2000 | 2.85700 | 2.75550 | -101.50 | 2.81800 | 2.73450 | -83.50 |

We thus present the responses reported in Fig. 7, showing the differential frequency (ΔF_S11) of each device after 10 minutes of the different concentrations (noted C). It increases for both devices according to the increase of the concentration of ethanol vapor, making it possible to verify a coherent behavior, associated with an effect on the sensitive material. From these curves, it can be estimated that the sensitivity is equal to 0.9 kHz/ppm and 1.3 kHz/ppm for D1 and D2, respectively for concentrations ranging from 500 to 1300 ppm.

IV. CONCLUSION

The feasibility of a fully inkjet printed capacitive microwave sensor for toxic vapor detection has been demonstrated, by showing that the ethanol vapor concentrations can be measured in real time. In addition, the sensitivity can be improved by increasing the sensitive layer thickness. In future work, the real time detection results will be shown and the sensitivity will be increased by further functionalizing this layer. This aims at enhancing the detection threshold and sensor resolution, in addition to addressing the selectivity issue which is commonly considered as a major concern in the chemical sensor field.

ACKNOWLEDGMENTS

The authors are grateful to the French National Research Agency for support under project ANR-13-BS03-0010 and the French RENATECH network (French National Nanofabrication Platform), as well as the French Embassy of Singapore (Merlion project).

REFERENCES

[1] Osha, Chemical Sampling Information – Ethyl Alcohol, 2017, https://www.osha.gov/dts/chemicalsampling/data/CH_239700.html (last consultation on15.02.17).
[2] RYDOSZ, Artur, et al. Microwave-based sensors with phthalocyanine films for acetone, ethanol and methanol detection. Sensors and Actuators B: Chemical, 2016, vol. 237, p. 876-886.
[3] A. Vena, L. Sydänheimo, M.M. Tentzeris, et al., A fully inkjet-printed wirelessand chipless sensor for $CO_2$ and temperature detection, IEEE Sens. J. 15 (1)(2015) 89–99, http://dx.doi.org/10.1109/JSEN.2014.2336838.
[4] BAHOUMINA, Prince, et al. Microwave flexible gas sensor based on polymer multi wall carbon nanotubes sensitive layer. Sensors and Actuators B: Chemical, 2017.
[5] PARAGUA, Carlos, et al. Study and characterization of CNT Inkjet printed patterns for paper-based RF components. In : 2015 European Microwave Conference (EuMC), IEEE, 2015, p. 861-864.